\documentclass[prd,12pt,a4paper,nofootinbib,showpacs,tightenlines]{revtex4}
\usepackage[left=2.5cm,right=2.5cm,top=2.5cm,bottom=2.5cm]{geometry}

\usepackage[frenchb,english]{babel}
\usepackage[latin1]{inputenc}
\usepackage{enumerate}
\usepackage{amsmath}
\usepackage{amsthm}
\usepackage{amssymb}
\usepackage{graphicx}
\usepackage{floatflt}
\usepackage{fancybox}
\usepackage{appendix}
\usepackage{enumitem}
\usepackage{subfigure}
\usepackage{array}

\usepackage{hyperref}
\usepackage[b]{esvect}

\usepackage{color}

\renewcommand{\Re}{\textrm{Re}}
\renewcommand{\Im}{\textrm{Im}}

\newcommand{\om}{\omega}
\newcommand{\Om}{\Omega}

\newcommand{\lam}{\lambda}

\newcommand{\Lam}{\Lambda}

\newcommand{\eff}{{\rm eff}}

\renewcommand{\u}{{u}}
\newcommand{\+}{{+}}
\renewcommand{\-}{{-}}
\renewcommand{\v}{{d}}

\newcommand{\M}{{\rm max}}

\newcommand{\conj}{{\psi}}
\newcommand{\p}{\partial}

\newcommand{\be} {\begin{equation}}
\newcommand{\ee} {\end{equation}}
\newcommand{\bsub}{\begin{subequations}}
\newcommand{\esub}{\end{subequations}}
\newcommand{\bea}{\begin{eqnarray}}
\newcommand{\eea}{\end{eqnarray}}
\newcommand{\bi} {\begin{itemize}}
\newcommand{\ei} {\end{itemize}}
\newcommand{\ben} {\begin{enumerate}}
\newcommand{\een} {\end{enumerate}}
\newcommand{\bmat} {\begin{pmatrix}}
\newcommand{\emat} {\end{pmatrix}}
\newcommand{\bal} {\begin{aligned}}
\newcommand{\eal} {\end{aligned}}
\newcommand{\btab}{\begin{tabular}}
\newcommand{\etab}{\end{tabular}}

\newcommand{\eq}[1]{equation~\eqref{#1}}

\usepackage{color}
\definecolor{myOrange}{rgb}{1,0.5,0.1}
\definecolor{myRed}{rgb}{0.8,0.1,0.1}
\definecolor{myGreen}{rgb}{0.7,0,0.8}
\definecolor{myGray}{rgb}{0.6,0.6,0.6}

\definecolor{light-gray}{gray}{0.95}

\begin{document}
\selectlanguage{english}

\title{Low frequency analogue Hawking radiation: \\ The Korteweg-de Vries model}

\author{Antonin Coutant}
\email{antonin.coutant@nottingham.ac.uk}
\affiliation{School of Mathematical Sciences, University of Nottingham, University Park, Nottingham, NG7 2RD, United Kingdom \\
Centre for the Mathematics and Theoretical Physics of Quantum Non-Equilibrium Systems,
University of Nottingham, Nottingham NG7 2RD, United Kingdom}
\author{Silke Weinfurtner}
\email{silke.weinfurtner@nottingham.ac.uk}
\affiliation{School of Mathematical Sciences, University of Nottingham, University Park, Nottingham, NG7 2RD, United Kingdom \\
Centre for the Mathematics and Theoretical Physics of Quantum Non-Equilibrium Systems,
University of Nottingham, Nottingham NG7 2RD, United Kingdom}

\begin{abstract}
We derive analytic expressions for the low-frequency properties of the analogue Hawking radiation in a general weak-dispersive medium. A thermal low-frequency part of the spectrum is expected even when dispersive effects become significant. We consider the two most common class of weak-dispersive media and investigate all possible anomalous scattering processes due inhomogeneous background flows. We first argue that under minimal assumptions, the scattering processes in near-critical flows are well described by a linearized Korteweg-de Vries equation. Within our theoretical model greybody factors are neglected, that is, the mode co-moving with the flow decouples from the other ones. We also exhibit a flow example with an exact expression for the effective temperature. We see that this temperature coincides with the Hawking one only when the dispersive length scale is much smaller than the flow gradient scale. We apply the same method in inhomogeneous flows without an analogue horizon. In this case, the spectrum coefficients decrease with decreasing frequencies. Our findings are in agreement with previous numerical works, generalizing their findings to arbitrary flow profiles. Our analytical expressions provide estimates to guide ongoing experimental efforts.
\end{abstract}


\date{\today}

\maketitle

%\newpage
%\tableofcontents

\section{Introduction}
In 1974 Hawking predicted the black hole evaporation process~\cite{Hawking74,Unruh76}, where a black hole spontaneously emit a thermal flux of radiation, thereby gradually reducing its mass. The black hole temperature is fully determined by a single local geometrical quantity, namely the surface gravity (noted $\kappa$) of the black hole at the event horizon. The surface gravity for an astrophysical (non-rotating and non-charged) black hole with mass M is $\kappa=1.010/M$ [kg/s], and the Hawking temperature $T_H = 1.2 \cdot 10^{-12} \kappa$ [K]. Since the surface gravity is inversely proportional to the mass of the black hole, the upper bound of the Hawking temperature is given by the smallest black holes observed in our universe, which are stellar black holes with a mass of the order of a few solar masses (noted $M_\odot$). The largest expected Hawking temperature is therefore of the order of $T_H = 6.0\cdot 10^{-8}{M_\odot}/{M}$ [K]. Consequently it is difficult, if not impossible, to verify the black hole evaporation process through observations. Despite this predicament the black hole evaporation process is one of the most, if not the most, studied and debated theoretical processes related to interplay between gravity and quantum field theory. This is largely due to the intriguing connection between Hawking's black hole temperature and Beckenstein's black hole entropy~\cite{Bekenstein73}, which is nowadays a central point for several approaches to quantum gravity. Any experimental evidence to support the black hole evaporation process is of great importance.

In 1981, Unruh~\cite{Unruh81} introduced a broad class of systems where perturbations propagate on an effective spacetime geometry, now generally known as analogue gravity systems~\cite{Barcelo05}. To be more precise they are analogues of classical and\,/\,or quantum field theory in effective curved spacetime geometries. Within these systems it is possible to set up analogues of black hole horizons (i.e.~when the propagation speed of waves is equal to that of the background flow), and as Unruh mentioned already in 1981, they should exhibit the Hawking radiation. In a flow, the surface gravity is given by the gradient of the difference between the background flow velocity $v$ and the propagation speed $c$ of the waves at the horizon, that is
\be \label{eq-surface-gravity}
T_H = \frac{\kappa}{2\pi} \doteq \frac1{2\pi} \p_x(v-c)\vert_{\rm horizon}.
\ee
Analogue black holes present the opportunity to experimentally test the Hawking mechanism. In recent years, several experiments were carried out that confirmed Unruh's prediction and obtained signatures of the analogue Hawking radiation. These were performed in optical systems~\cite{Belgiorno10}, open channel flows~\cite{Weinfurtner10,Euve15} and Bose-Einstein condensates~\cite{Steinhauer14,Steinhauer16}.

However, in general in an analogue system, the emitted spectrum differs from a thermal one at the Hawking temperature \eqref{eq-surface-gravity}. In this work we present a theoretical analysis that investigates in depth the \emph{deviations} from the predicted Hawking temperature \eqref{eq-surface-gravity}. The main source of deviations comes from the dispersive nature of the media, which introduces higher order spatial derivatives. Because of this, analogue gravity systems are a natural test bed to investigate the universality and robustness of the black hole evaporation process against short-distance modifications. Our concern here is to investigate the influence of dispersion in the low-frequency part of the spectrum. This is of great interest for ongoing experimental~\cite{Weinfurtner10,Euve15,Belgiorno10,Steinhauer14,Steinhauer16} and numerical~\cite{Corley96,Macher09b,Finazzi10b,Finazzi11,Finazzi12,Michel14,Michel15,Robertson16} efforts to investigate the Hawking spectra, as the low-frequency behaviour is accessible.

There is in principle another source of deviation which affects the thermality of the black hole spectra in an analogue gravity system. This other source is greybody factors, which are expected to occur in both analogue~\cite{Anderson14,Anderson15,Fabbri16} and astrophysical~\cite{Page76} black holes. Greybody factors account for the partial reflection of the Hawking radiation propagating away from the horizon back into the black hole. They are equivalent to the Albedo-coefficients of a black body radiation. Since greybody factors are also expected to modify the low-frequency regime of the Hawking spectra, it is difficult in principle to disentangle the two sources of deviation. One approach commonly used to describe the propagation of surface waves in open channel flows is the Korteweg-de Vries (KdV) model. We argue here, and demonstrate in a companion paper~\cite{Coutant17b} where we include greybody factors, that the KdV model is giving an accurate description of the analogue Hawking effect for near-critical flows. Furthermore, we show that the KdV model can be applied to all known analogue gravity systems in the weak-dispersive regime. To do this we consider two types of dispersion relations, whether the propagation speed decreases at short distances, as for example for sound and surface waves in classical fluids, or increases at short distances, as in Bose-Einstein condensates. As we will explain in detail below, the KdV model in both cases neglects greybody factors, and one can calculate the correction to the Hawking spectra only due to dispersion for all scattering processes exhibiting anomalous wave scattering.

\section{Linear Korteweg-de Vries model}
To derive the Hawking effect in General Relativity, one builds the quantum state of the radiation field by tracing back the modes to the formation of the black hole~\cite{Hawking74}, or to a region close to the horizon~\cite{Unruh76}, where physical state (the Unruh vacuum) can be constructed. During this back tracing process, the modes undergo an exponential blueshift. For this reason, the derivation of the Hawking effect seems to rely on the physics of ultra high energy modes, where the notion of a classical and smooth space-time becomes dubious. This is referred to as the ``Transplanckian problem''~\cite{Jacobson96,Jacobson99}. In analogue systems, one must take into account the deviation of the linear dispersion relation, which inevitably arise at short wavelengths~\cite{Jacobson91}. For this reason, we now consider the propagation of wave in a weak dispersive regime. In this section we look at plane waves in a homogenous fluid flow.

\subsection{Dispersive plane waves on a flow}
The frequency $\om > 0$ and wave number $k$ of a plane wave $\phi = \Re(A e^{-i \om t + i k x})$ propagating in a homogenous fluid are related by the dispersion relation of the medium $\om^2 = F(k)$. If the flow velocity of the fluid is non-zero, the dispersion relation is modified by the Doppler effect. We will consider waves propagating on an effective one-dimensional fluid flow aligned with the $x$-axis of the coordinate systems such that $v>0$. The frequency $\Om$ as measured in the fluid frame (Fig.~\ref{Frame_Fig}(a)) obeys the dispersion relation, but the frequency in the lab frame (Fig.~\ref{Frame_Fig}(b)) is related by the Doppler relation $\Om = \om - v k$. Therefore, the dispersion relation takes the general form
\be \label{HJ_general}
(\om - v k)^2 = F(k).
\ee
This relation describes the propagation in a variety of media. For instance, the dispersion relation for surface gravity waves is obtained by choosing~\cite{Johnson,Mei}
\be \label{Eq-DispRel-GravityWaves}
F_\mathrm{gw}(k)=\left(gk + \frac{\sigma}{\rho} k^3  \right) \, \tanh(k h),
\ee
where $g$ is the local gravitational acceleration on earth (hence, the name gravity waves), $\sigma$ the surface tension, $\rho$ the water density and $h$ the water height. In Fig.~\ref{Frame_Fig} we depict the dispersion relation of gravity waves for a homogenous sub-critical fluid flow $v^2<c^2$. For sound waves in Bose-Einstein condensates one has to pick~\cite{Pitaevskii}
\be \label{Eq-DispRel-BEC}
F_\mathrm{BEC}(k)=c^2 k^2 + \frac{\hbar^2}{4 m^2}k^4,
\ee
where $c$ is the propagation speed of the sound waves, $\hbar$ is the reduced Planck constant and $m$ the mass of the fundamental Bosons underlying the superfluid.

These two dispersion relations have been extensively discussed in analogue gravity when discussing the analogue Hawking effect. Both gravity wave~\cite{Weinfurtner10,Euve15} and Bose-Einstein condensate~\cite{Steinhauer15,Steinhauer14} based analogue gravity experiments are currently the most successful candidates to attest for the robustness and universality of black hole effects in the lab. Besides practical reasons, they are an example for each super-critical (\eq{Eq-DispRel-BEC}) and sub-critical (\eq{Eq-DispRel-GravityWaves}) dispersion relations. To make the comparison between the two more apparent, it is instructive to consider surface waves in the so-called ``weak dispersive regime,'' where $F_\mathrm{gw}(k)$ is Taylor expanded up to fourth order in $k$:
\be \label{F-GW-wd}
F_\mathrm{gw}(k)\sim g h k^2-\left(\frac{g h^3}{3} - \frac{h \sigma}{\rho}\right) k^4.
\ee
Within the weak-dispersive regime both dispersion relations (\ref{Eq-DispRel-BEC}) and (\ref{F-GW-wd}) read
\be \label{F-disp}
F_\mathrm{disp} = c^2 k^2 \pm \frac{k^4}{\Lambda^2},
\ee
where $\Lambda$ is a constant, characteristic of dispersive effects. When the sign of the last term in \eqref{F-disp} is positive, dispersive effects increase the propagation speed (we refer to this case as `superluminal'), while if it is negative, the propagation speed is decreased (`subluminal' case). Hence, in BECs $\Lambda_\mathrm{BEC} = 2m/\hbar$ and the dispersion is superluminal, while for gravity waves $\Lambda_\mathrm{GW}=(h (gh^2/3-\sigma/\rho))^{-1/2}$ and the dispersion is subluminal. As a side remark it is possible to tune $\Lambda_\mathrm{GW}^{-1} = 0$, by choosing the water height $h$ accordingly.

\begingroup
\centering
\begin{figure}[htbp]
\centering
\subfigure[\label{fig:1a}$\;$\textbf{Fluid frame}]{\includegraphics{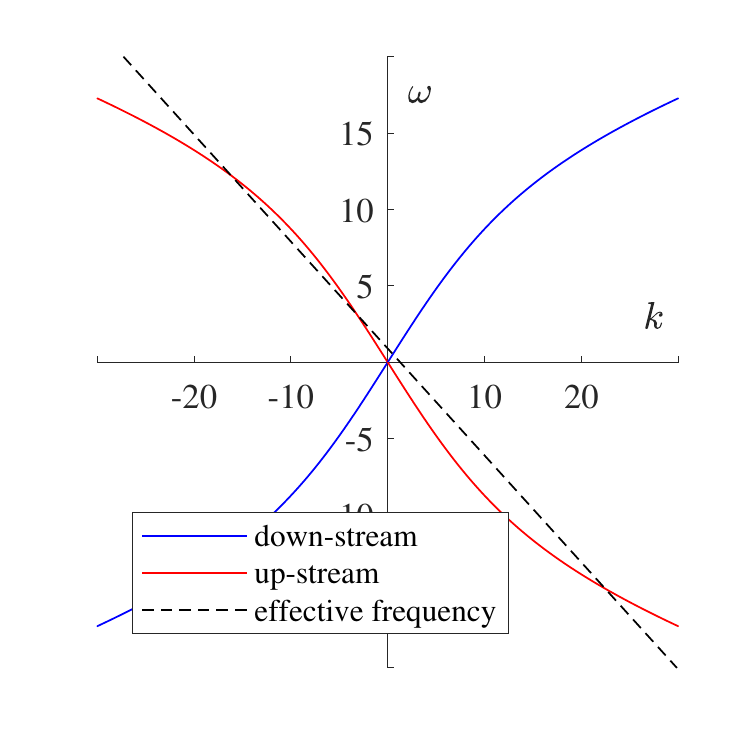}}
\subfigure[\label{fig:1b}$\;$\textbf{Laboratory frame}]{\includegraphics{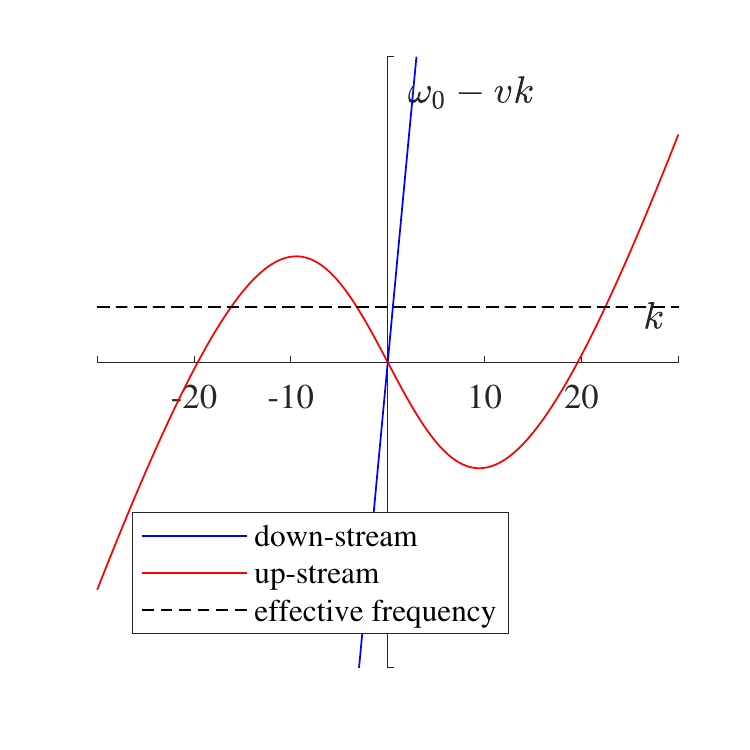}}
\caption{\label{Frame_Fig}\textbf{Fluid versus laboratory frame}. We plot the dispersion relation $\omega_\eff = \pm \sqrt{F_\mathrm{gw}}$ in the fluid co-moving frame (a) and in the laboratory or rest frame (b). Within each frame we divide the branches of the dispersion relation, such that plane waves with wave number $k$ are either co-moving (blue lines) or counter-moving (red lines) with respect to the fluid flow $v>0$. In both frames we also illustrate the permitted $k$-values for a single frequency mode with $\om_0>0$, which are on the intersections of the red\,/\,blue with the black dashed lines. In the fluid frame we plot the effective frequency $\om_\mathrm{eff}=\om_0-vk$ (black dashed line in panel a), while in the laboratory frame the frequency of the mode remains constant $\om_\eff = \om_0$ (b). (We used the following parameters for both plots $\sigma=0$, $g=1$, $c=1$, $h=1$, $\om_0=0.1$ and $v=0.7$ in standard units.)}
\end{figure}

\endgroup

The next step is to introduce the linear Korteweg-de Vries (KdV) model, which will allow us to investigate the modifications of the Hawking spectra exclusively due to dispersive effects. To proceed, we split the set of solutions into two classes: one corresponds to solutions moving against the flow, while the other describes solutions moving with the flow (we refer to appendix~\ref{KdV_App} for more details). In Fig.~\ref{Frame_Fig} the modes moving against\,/\,with the flow are depicted red\,/\,blue. Modes that propagate with the flow are simply absorbed by the analogue black hole, and they are not at the origin of the Hawking radiation. However, in general, these modes couple with the Hawking modes, and dress the outgoing flux, giving rise to what is called greybody factors in the black hole literature~\cite{Page76}. In this work, by working with the KdV approximation, we neglect this coupling, in order to focus on the effects due to dispersion.

To split the co- and counter-propagating sectors, we take the two square roots of \eq{HJ_general}. Moreover, since we work in the weak dispersive regime, we only keep the first dispersive correction. In other words, we Taylor-expand $\sqrt{F_\mathrm{disp}}$ in $k$ (\ref{F-disp}) to next-to-leading order. This gives us
\bsub  \bea
\om_{\rm sub} &=& (v \pm c) k \mp \frac{k^3}{\lam^2}, \label{HJ-sub} \\
\om_{\rm super} &=& (v \pm c) k \pm \frac{k^3}{\lam^2},
\eea \esub
where $\lam^2=2c\Lambda^{2}$ is introduced to enlighten the notations. The subscript in the above equation refers to the type of dispersion relation, as in \eqref{F-disp}. Note that the signs for the sub-\,/\,super-critical dispersive media in front of the dispersive term are interchanged, such that the counter-propagating modes $\tilde{\om}$ in a sub-critical dispersive media in KdV approximation are described by
\bsub \label{KdVdisprel} \bea
\tilde{\om}_{\rm sub} &=& (v - c) k + \frac{k^3}{\lam^2}, \\
\tilde{\om}_{\rm super} &=& (v - c)k -  \frac{k^3}{\lam^2},
\eea \esub
As we will point out below in section~\ref{BHWH_Smat_Sec} it is sufficient to investigate one of the two dispersive media, as the other will follow due to symmetry arguments. We choose to focus on the case of water waves (sub-critical dispersive media). In Fig.~\ref{HJ_Fig}, we plotted both branches of the dispersion relation~\eqref{HJ-sub} in both homogenous super-critical~\ref{HJ_Fig_2a} and sub-critical~\ref{HJ_Fig_2b} flows in the laboratory frame. In Fig.~\ref{Fig_compare_theory}, we compare the KdV approximation with the full dispersion relation of water waves in equation \eqref{Eq-DispRel-GravityWaves}. 

\begingroup %\label{HJ_Fig}
%\centering %[width=0.45\textwidth]
\begin{figure}[htbp]
\centering
\subfigure[\label{HJ_Fig_2a}$\;$\textbf{$\mathbf{v/c=1.15}$}]{\includegraphics{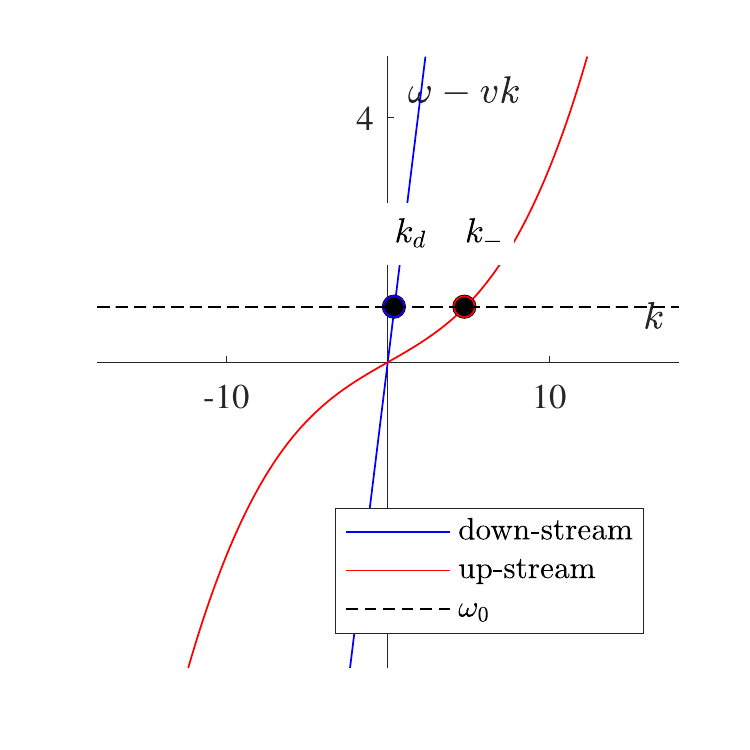}}
\subfigure[\label{HJ_Fig_2b}$\;$\textbf{$\mathbf{v/c=0.7}$}]{\includegraphics{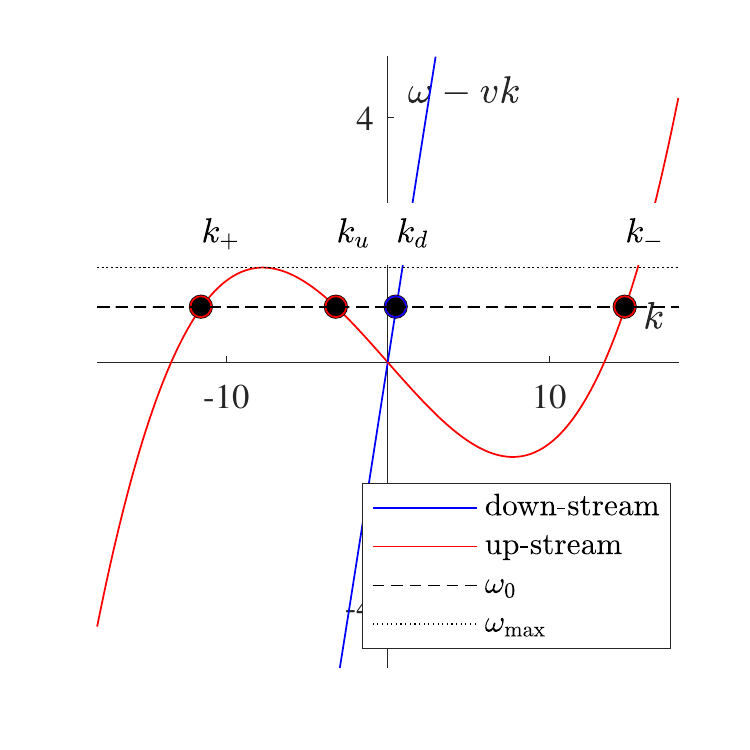}}
\caption{\label{HJ_Fig}\textbf{Sub-critical dispersion relation in both super- and sub-critical fluid flows}.
We plot the down-stream (all solid blue lines) and the up-stream (all solid red lines) branches of the KdV dispersion relation for surface waves (sub-critical dispersive media). In panel (a) we consider a homogenous super-cricital flow $v/c=1.15$, while, in panel (b) we investigate a homogenous super-critical flow $v/c=0.7$. For a fixed frequency $\om_0<\om_\M$ we have two\,/\,four solutions indicated by the intersections between the dashed black lines with the dispersion relations branches in the super-\,/\,sub-critical flow. For $\om > \om_\M$, there are two solutions, on both super-\,/\,sub-critical flow. The latter case will not be considered in this paper, since we focus on low frequencies.}
\end{figure}

Before discussing the dynamics of waves, we first discuss the solutions of \eq{HJ-sub} for both a sub-critical flow $v < c$, and a super-critical one $v > c$ (see Fig.~\ref{HJ_Fig}). When the flow is sub-critical (panel b), and below a certain frequency $0 < \om < \om_\M$, there are 3 solutions $\left\{k_{\+},k_\u,k_\- \right\}$ in the counter-propagating branch of the dispersion relation (solid red line). The wave number $k_{\u}$ describes a long wavelength, non-dispersive wave propagating against the flow. It is non-dispersive in the sense that it exists irrespectively of the value of $\lam$. On the other hand, the two other solutions $\left\{k_{\+},k_\- \right\}$ disappear in the limit $\lam \to \infty$. For this reason, we shall refer to them as the dispersive roots. The peculiarity of the Hawking problem is that one of the dispersive roots has a negative norm, or equivalently, a negative energy (see below). This means that when this mode is excited, the total system (background flow plus excitations) has a lower energy than the background flow alone. To make this explicit, the subscript $\pm$ of $k_\pm$ refers to the sign of the norm of the corresponding mode.

\begingroup
%\centering %[width=0.45\textwidth]
\begin{figure}[htbp]
\centering
\subfigure[\label{fig:3a}$\;$\textbf{$\mathbf{v/c=0.5}$}]{\includegraphics[width=0.32\textwidth]{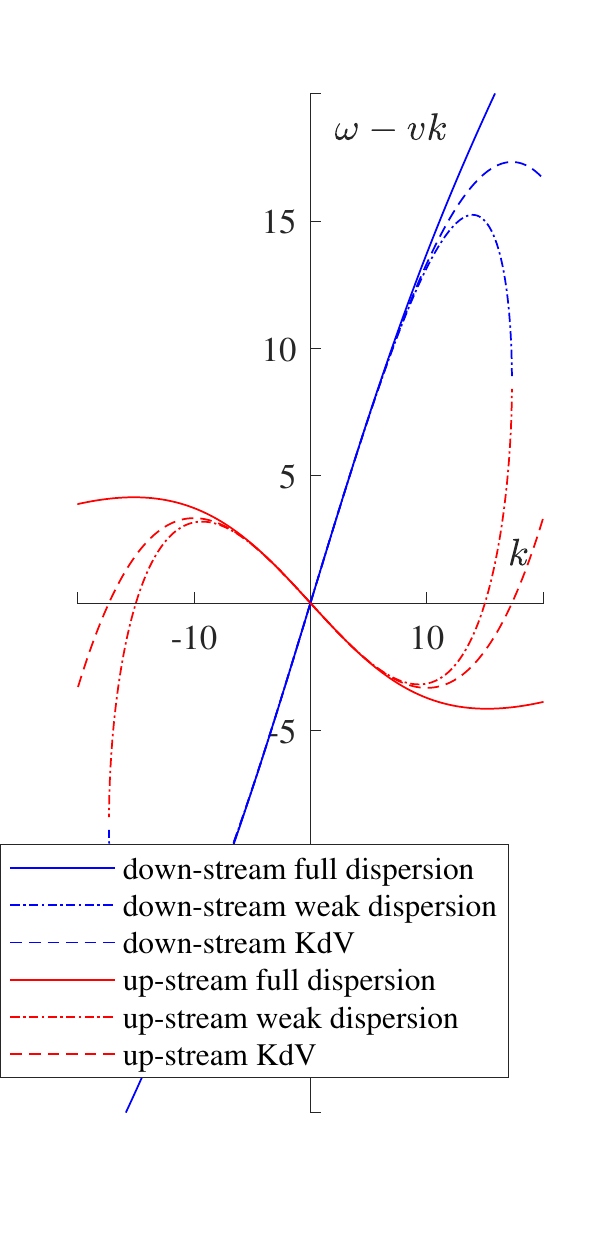}}
\subfigure[\label{fig:3b}$\;$\textbf{$\mathbf{v/c=0.9}$}]{\includegraphics[width=0.32\textwidth]{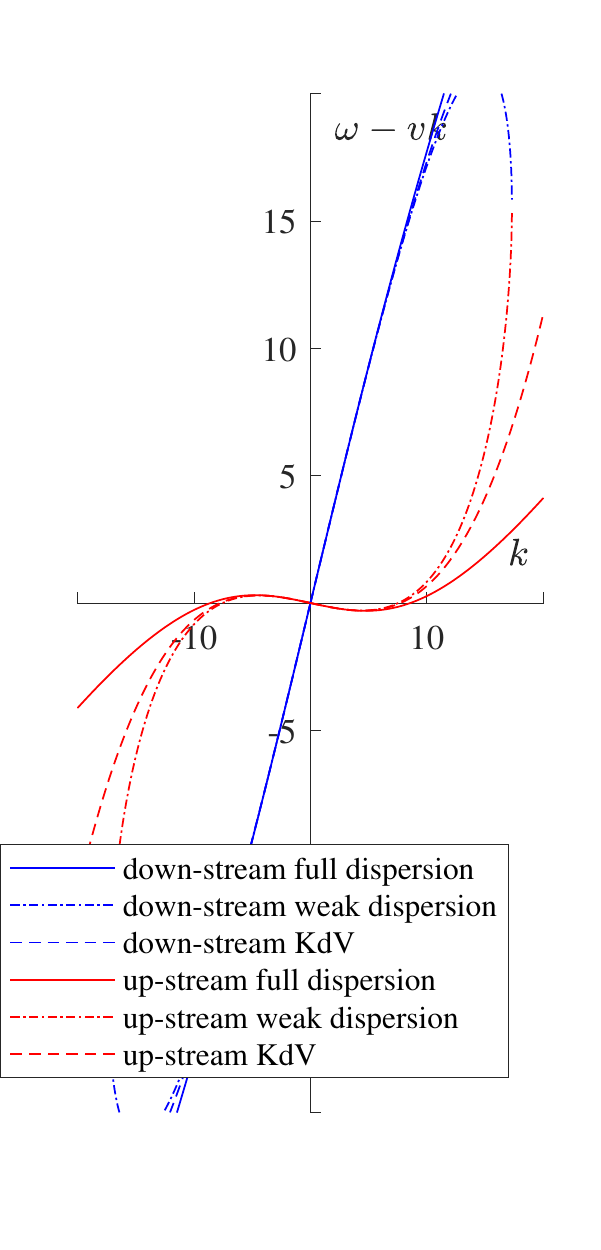}}
\subfigure[\label{fig:3c}$\;$\textbf{$\mathbf{v/c=1.2}$}]{\includegraphics[width=0.32\textwidth]{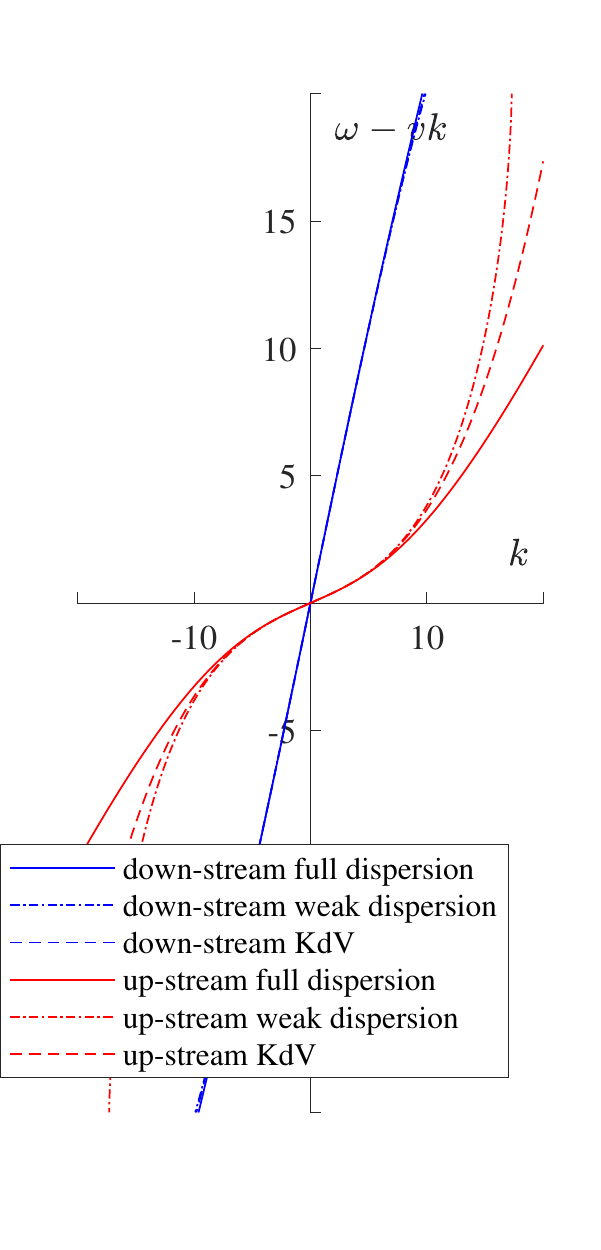}}
\caption{\label{Fig_compare_theory}\textbf{Comparison of various approximations to describe surface gravity waves}. We plot the down-stream (all blue lines) and the up-stream (all red lines) branches of the dispersion relation $\omega-vk=\sqrt{F}$ for surface waves in the various levels of approximations discussed here. In the absence of surface tension ($\sigma=0$), we plot the full (solid red and blue lines), the weak-dispersive (dotted red and blue lines), and the KdV (dashed red and blue lines) dispersion relations. In panel (a) we consider a very slow flow $v/c=0.5$, while, in panel (b) we investigate close-to-critical flows $v/c=1$ and in panel (c) we consider a super-critical flow. We are interested in the flow-frequency limit and the applicability of KdV approximation. It is illustrated that the KdV approximations is a better solution in close-to-critical and critical flows, as the lines of the three different approximations converge as $\om\rightarrow 0$.}
\end{figure}

\subsection{Equation of motion}
To investigate the Hawking process in any analogue gravity system it is necessary to consider the case of waves propagating on a non-homogeneous flow. In particular, in an analogue black or white hole, the background quantities $v(x)$, $c(x)$ depend on the position, and there is a transition region where the flow passes from super-critical ($c^2<v^2$) to sub-critical ($c^2>v^2$). The point where $v^2 = c^2$ is the analogue horizon.

Since we exclude the coupling to the co-propagating wave, we assume that counter-propagating waves are solutions of the linear Korteweg-de Vries equation. For this, we start from the action
\be \label{KdVaction}
S = \frac12 \int \left[ \p_t \phi \p_x \phi - (c-v) (\p_x \phi)^2 \pm \frac{(\p_x^2 \phi)^2}{\lam^2} \right] dt dx,
\ee
which indirectly implies the KdV as the equation of motion
\be
- \p_t \p_x \phi + \p_x (c-v) \p_x \phi \pm \frac{\p_x^4 \phi}{\lam^2} = 0.
\ee
Indirectly, because $\phi$ is the variable in the action, but it is its conjugate momentum $\conj = \p_x \phi$ that obeys the Korteweg-de Vries equation in its standard form. As a consequence one of the solutions of this equation is a constant in space. As a side remark, we point out that this constant solution can be seen as a relic of the down-stream mode $\phi_\v$. In order to eliminate it in the equation of motion, we integrate the above equation and fix the integration constant to 0. At fixed frequency $\om$, we obtain
\be \label{KdVmodeq}
i \om \phi + (c-v) \p_x \phi \pm \frac{\p_x^3 \phi}{\lam^2} = 0.
\ee

%{Equation of motion in the low-frequency limit}
Our aim is now to solve the scattering problem in the low-frequency limit $\om \to 0$. For this, we first set $\om=0$ in the equation of motion. It results that $\p_x \phi = \psi$ satisfies a second order ordinary differential equation, given by
\be \label{SchroEq}
\pm \p_x^2 \conj = - \lam^2(c - v) \conj  .
\ee
To obtain the scattering coefficients we proceed in two steps. We first solve \eqref{SchroEq} to obtain the solutions of the mode equation \eqref{KdVmodeq} at $\om = 0$. Second, we obtain their asymptotic behaviors for $x \to \pm \infty$, and identify them to the $\om \to 0$ expansion of superposition of plane waves.

\subsection{Conserved quantities}
Before solving this equation, we exploit the fact that it comes from an action, and, hence possesses several conserved quantities. The first one is the canonical norm, which reads
\be \label{norm}
N(\phi,\phi) = \Im \int \phi^* \p_x \phi dx.
\ee
The second one is the conserved energy of a wave, which is given by
\be
E(\phi,\phi) = \int \p_t \phi \p_x \phi dx.
\ee
For a single frequency mode $\phi = 2\Re(\phi_\om(x) e^{-i \om t})$, with $\om > 0$, the energy is simply given by the frequency times its norm $E_\om = \om N_\om$. Hence, the negativity of the norm is equivalent to that of the energy (we refer to~\cite{Coutant13} for a careful discussion of the link between norm and energy). We also point out that when working in the ray approximation, the canonical norm reduces to the wave action~\cite{Buhler}, as we see from the relation $N_\om = E_\om/\om$. In the following, we will work using a normalized basis of modes, that is, such that $N(\phi_\om,\phi_{\om'}) = \pm \delta(\om - \om')$. Any solution $\phi_\om$ of the mode \eq{KdVmodeq} can be written as a superposition of normalized plane waves, given by
\be \label{eqn-norm-decomp-pw}
\phi_j = \frac{e^{i k_j x}}{\sqrt{\left|k_j v_g\right|}} ,
\ee
where $k_j$ is one of the solutions of the dispersion relation, and
\be
v_g = \left. \frac{\partial\om}{\partial k} \right\vert_{k_j}=-(c-v)\pm\frac{3k_j^2}{\lambda^2}
\ee
the group velocity of the mode, compare also with~\cite{Richartz12}.

\section{Scattering matrix in the KdV model\label{Smat_Sec}}
We are now ready to tackle the full problem of inhomogeneous flows. For this we assume that the flow accelerates or decelerates over a region of size $\approx L$. When $x \ll -L$ or $x \gg L$, the flow is constant, that is
\be \label{Vas}
v(x) \to v_{r/l} \quad (x \to \pm \infty), \qquad \textrm{and} \qquad c(x) \to c_{r/l} \quad (x \to \pm \infty),
\ee
and, hence the solutions are given by superpositions of plane waves. The linear relation between modes going toward the transition (as read from the group velocity $v_g$) and modes going away from it defines the scattering matrix ($S$-matrix). An analogue white hole (resp. black hole) is obtained when the flow velocity $v$ crosses the speed of waves $c$ by decelerating (resp. accelerating). We start by considering the white hole case, and as we shall see, the black hole case directly follows. Since $v>0$ by assumption, this means that the flow is supercritical on the left side ($v_l > c_l$), and subcritical on the right side ($v_r < c_r$).

Although not the subject of the current study, we would like to initiate the discussion on the $S$-matrix for weak-dispersive media beyond the KdV approximation, see equation~\eqref{F-disp}. This is to show in detail how the scattering process is is simplified in the KdV approximation. (Note that in our second paper in this series we will discuss this  in great detail, see~\cite{Coutant17b}.) For sub-critical weak-dispersive media there are 2 asymptotic modes on the super-critical side and 4 on the sub-critical one; for super-critical weak-dispersive media there are 4 asymptotic modes on the super-critical side and 2 on the sub-critial one.
Since we are interested in low frequencies, we always assume $\om < \om_\M$ (see Fig.~\ref{HJ_Fig}). In a white hole flow, after identifying the \emph{in} and \emph{out} modes~\footnote{To ease the identification of \emph{in} and \emph{out} modes, we quickly discuss the group velocities of the various modes of Fig.~\ref{HJ_Fig}. Both dispersive roots $k_\pm$ propagate to the left. On the left side the flow is super-critical, hence $k_-$ is the only root. It propagates towards the horizon, and, hence is \emph{in}-going. On the right, the flow is sub-critical. $k_\pm$ are both propagating away from the horizon, while $k_\u$ propagates towards it.}, the $S$-matrix is defined by
\be \label{fullSmat}
\bmat \phi_\u^{\rm in} \\ \phi_\-^{\rm in} \\ \phi_\v^{\rm in} \emat = S_{\rm WH} \cdot \bmat \phi_\+^{\rm out} \\ \phi_\-^{\rm out} \\ \phi_\v^{\rm out} \emat =
\bmat \alpha & \beta & R \\ \tilde \beta & \tilde \alpha & B \\ \tilde R & \tilde B & \tilde T \emat \cdot \bmat \phi_\+^{\rm out} \\ \phi_\-^{\rm out} \\ \phi_\v^{\rm out} \emat.
\ee
Because we work with normalized modes the conservation of the norm~\eqref{norm} (or energy) implies that $S \in U(2,1)$. This gives several relations between the coefficients. For instance
\be
|\alpha|^2 - |\beta|^2 + |R|^2 = 1.
\ee
We now consider the $S$-matrix for weak-dispersive media in the KdV approximation.

\subsection{Scattering matrix for transitions from sub- to super-critical flows and vice versa \label{BHWH_Smat_Sec}}
In the KdV approximation, we neglect $\phi_\v$, as this modes decouples completely from the other modes. This means that we have $R = B = \tilde R = \tilde B = 0$ in \eq{fullSmat}. Discarding the down-stream mode, the three others are described by the solution of the KdV dispersion relation \eqref{KdVdisprel}. Then, for sub-critical media in the KdV approximation there are 1 asymptotic modes on the super-critical side and 3 on the sub-critical one; for super-critical media in the KdV approximation there are 3 asymptotic modes on the super-critical side and 1 on the sub-critical one.

Since we have one less root on each side of the transition area, the $S$-matrix reduces to a simpler form, element of $SU(1,1)$, given by
\be
S_{\rm WH} = \bmat \alpha & \beta \\ \tilde \beta & \tilde \alpha \emat,
\ee
and the relation between the coefficients reduces to
\be
|\alpha|^2 - |\beta|^2 = |\tilde \alpha|^2 - |\tilde \beta|^2 = 1.
\ee
In principle, there are 4 different scattering processes one can investigate in our setup: scattering on a (a) white hole horizon in a sub-critical dispersive media; (b) black hole horizon in a sub-critical media; (c) white hole horizon in a super-critical dispersive media; and last but not least (d) black hole horizon in a super-critical dispersive media. As already mentioned above, one only needs to calculate one of the four cases to obtain the three others using symmetry arguments. In appendix~\ref{KdV_App} we show the relations between the $S$-matrices of the 4 possible transcritical flows (this is a generalization of~\cite{Coutant11}). Explicitly, we have:
\be \label{S_sym}
S_{\rm WH}^- = (S_{\rm BH}^-)^{-1} = (S_{\rm BH}^+)_{l \leftrightarrow r}^{-1} = (S_{\rm WH}^+)_{l \leftrightarrow r}.
\ee
The superscrit $+$ or $-$ indicates the type of dispersion relation, i.e. the sign in \eq{F-disp} (and, hence in \eqref{KdVmodeq}), and the subscript $l \leftrightarrow r$ indicates that one swaps the role of left and right sides. As a last remark, we point out that these correspondences work when the approximations of the KdV model hold. In particular, greybody factors differ in general in these 4 cases.

In the recent experimental literature there are in particular 2 cases that have been investigated in depth: (a) the scattering of white hole horizons in sub-critical  media~\cite{Weinfurtner10} and (d) the scattering on black hole horizons in super-critical media~\cite{Steinhauer15}. Since the two cases are related by \eqref{S_sym}, we will calculate case (a) and by applying the transformation \eqref{S_sym} to the resulting scattering coefficients we automatically cover (d).

\subsection{Scattering matrix for inhomogeneous sub- or super-critical flows}
We now consider transitions in a flow which remains subcritical or supercritical throughout within the KdV model. If dispersive modes are present, there will be a mode-mixing between positive and negative norm modes, see for example~\cite{Finazzi11,Coutant16}. This happens in two cases: for (a) sub-critical flows with a sub-critical dispersion relation; (b) super-critical flows with a super-critical dispersion relation. Again, using the symmetry arguments as we explained in appendix~\ref{KdV_App}, we can focus on (a), and (b) is then obtained by symmetry.

We thus consider a flow that accelerates to the left, similarly to a white hole, but don't reach criticality, i.e. $0 < c_l - v_l < c_r - v_r$. Since the three modes now exist on both sides of the transition, the $S$-matrix is $3 \times 3$. It is defined as
\be \label{SubcritSmat}
\bmat \phi_\u^{\rm in} \\ \phi_\-^{\rm in} \\ \phi_\+^{\rm in} \emat = S \cdot \bmat \phi_\u^{\rm out} \\ \phi_\-^{\rm out} \\ \phi_\v^{\rm out} \emat =
\bmat T & \beta & \alpha \\ \tilde \beta & \tilde A & \tilde B \\ \tilde \alpha & B & A \emat \cdot \bmat \phi_\u^{\rm out} \\ \phi_\-^{\rm out} \\ \phi_\+^{\rm out} \emat
\ee
Conservation of the norm implies that $S \in SU(2,1)$.

This kind of scattering processes would happen for surface waves in inhomogeneous sub-critical flows~\cite{Weinfurtner10,Weinfurtner13,Euve14,Euve15,Michel15,Robertson16}, but also for sound waves in BECs in inhomogeneous super-sonic flows.

%%%%%%%%%%%%%%%%%%%%%%%%%%%%%%%%%%%%%%%%%%%%%%%%%%%
%KdV MODEL
%%%%%%%%%%%%%%%%%%%%%%%%%%%%%%%%%%%%%%%%%%%%%%%%%%%

\section{Scattering coefficients in the low-frequency limit}
We are now ready to solve \eq{KdVmodeq} for the scattering coefficients $\alpha$ and $\beta$ for a transition from a supercritical ($x\rightarrow -\infty$) to a subcritical flow ($x\rightarrow +\infty$) in a subcritical dispersive media in the low-frequency limit. We first derive general results, independent of the details of the flow profile, which allow us to obtain the $\omega$-dependence of the scattering coefficients. We then assume a specific profile to fully compute for the scattering coefficients in the low-frequency limit.

Since we are now interested in the solutions of \eq{SchroEq}, it is natural to define two adimensional wavenumbers, using the $\om = 0$ solutions of the dispersion relation \eqref{KdVdisprel} and the size of the transition $L$. Hence, we define
\bsub \bea
q_l &=& \lam L \sqrt{|c_l - v_l|} , \\
q_r &=& \lam L \sqrt{|c_r - v_r|} .
\eea \esub
In the rest of this section, we specifically focus on a sub-critical dispersion relation, the other case being deduced from \eq{S_sym}.

\subsection{Transcritical flows}
In transcritical flows, since $v_l > c_l$, the solutions of \eq{SchroEq} grow or decay exponentially for $x \to -\infty$. Since scattering modes must stay bounded in space, the only physically acceptable (up to a multiplicative constant) solution decays on the left side.

\subsubsection{General results}
In full generality, its asymptotic behavior is given by
\bsub
\bea
\conj &\sim& e^{q_l x/L} \quad (x \to -\infty), \qquad \mathrm{and}  \\
\conj &\sim& A_2 e^{i q_r x/L} + A_3 e^{-i q_r x/L} \quad (x \to +\infty).
\eea
\esub
From this we can obtain the two \emph{in} modes of a white hole flow. The long wavelength mode coming from the represented in Fig.~\ref{Corley_Fig}, decays exponentially on the left. Hence it is simply given by $\phi = \int_{-\infty}^x \conj(x') dx'$. It asymptotically behaves as
\bsub \label{Corley_coef} \bea
\phi &\sim& \frac{e^{q_l x/L}}{q_l/L} \quad (x \to -\infty), \qquad \mathrm{and} \\
\phi &\sim& A_1 + A_2 \frac{e^{i q_r x/L}}{i q_r/L} + A_3 \frac{e^{-i q_r x/L}}{-i q_r/L} \quad (x \to +\infty) .
\eea \esub
It is important to notice that the coefficients $A_{1,2,3}$ are independent of $\om$, since they are obtained from \eq{SchroEq} where $\om$ does not appear.

\begin{figure}[!ht]
\begin{center}
\includegraphics[width=0.8\columnwidth]{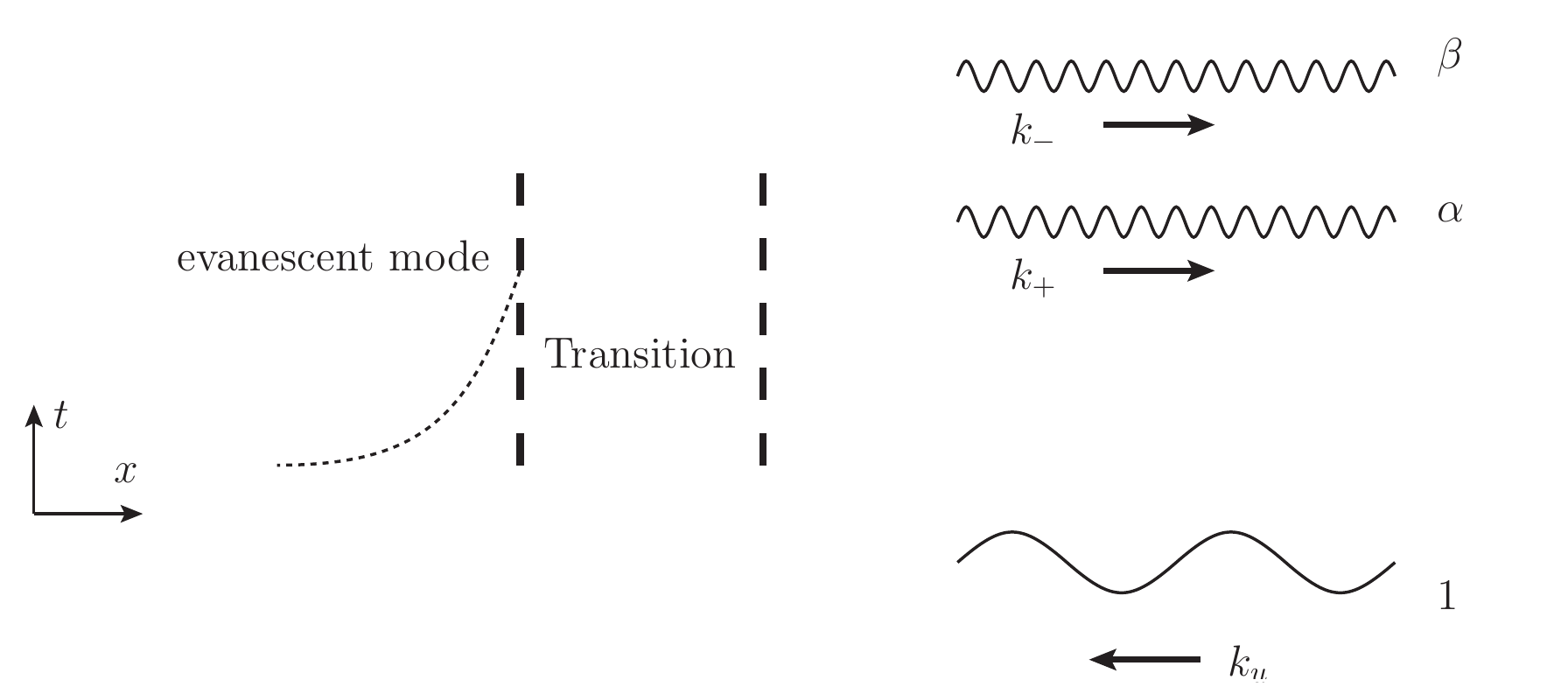}
\end{center}
\caption{\textbf{Schematic representation of scattering process related to an effective white hole horizon.} We consider the scattering process for a sub-critical dispersive media. We consider ingoing modes from the sub-critical side of the flow ($x\rightarrow +\infty$), hence we investigate the scattering of an effective white hole horizon. In the super-critical region of the flow, which we assume to be on the left ($x\rightarrow -\infty$) $\phi_\u^{\rm in}$ of the transition region, we only have a decaying\,/\,bound mode. On the sub-critical side far from the transition region (where the flow is homogenous again) we can decompose the ingoing field onto a basis of three plane waves.}
\label{Corley_Fig}
\end{figure}

The above equations~\eqref{Corley_coef} are valid far enough from the transition, that is, for $|x| \gg L$. Moreover, we know that in this region, solutions are given in terms of a superposition of plane waves, see equation~\eqref{eqn-norm-decomp-pw}. To identify such a superposition with the asymptotic of \eq{Corley_coef}, we use the $\om \to 0$ limit of these plane wave.
The incoming long wavelength mode of Fig.~\ref{Corley_Fig} is, for $|x| \ll c/\om$, given by
\be \label{Corley_mode}
\phi_\u^{\rm in} \sim \frac1{\sqrt{\om}} + \alpha \frac{e^{-i \lam \sqrt{2(c_r - v_r)} x}}{\sqrt{2 (c_r - v_r) q_r/L}} + \beta \frac{e^{i \lam \sqrt{2(c_r - v_r)} x}}{\sqrt{2(c_r - v_r) q_r/L}} \qquad (x \to +\infty).
\ee
This mode is schematically represented in Fig.~\ref{Corley_Fig}. We now identify equations \eqref{Corley_coef} and \eqref{Corley_mode} to extract the scattering coefficients. First we multiply our solution $\phi$ with the appropriate constant to normalize the incoming part as in \eqref{Corley_mode}, and then we obtain the scattering coefficients
\bsub \bea
\alpha &=& i \frac{A_3}{A_1} \sqrt{\frac{2 (c_r - v_r) L}{\om q_r}} , \\
\beta &=& -i \frac{A_2}{A_1} \sqrt{\frac{2 (c_r - v_r) L}{\om q_r}} \label{eq-beta-general} .
\eea \esub
What we have shown here, is that for any profile, the low-frequency behavior of the Bogoliubov coefficients is $|\alpha|^2 \sim |\beta|^2 \sim O(\om^{-1})$.
\bigskip

\textbf{Effective temperature in the low-frequency limit.}
The analogue Hawking radiation in the absence of dispersive effects (and neglecting grey-body factors) is given by equation~\eqref{eq-surface-gravity}. In the presence of dispersive effects but when they are negligible, that is when $\kappa \ll \lam \sqrt{c}$, we have $|\beta|^2 \sim T_H/\om$. The fact that $\omega \vert \beta \vert^2$ is constant in $\omega$ can be interpreted as equipartition of energy.  Since we have shown that $|\beta|^2 = O(1/\om)$ for any $\lam \sqrt{c}/\kappa$, it is natural to define an effective temperature as
\be
T_\eff \doteq \lim_{\om \to 0} \left(\om |\beta_\om|^2 \right) .
\ee
This effective temperature conveniently parametrize the Bogoliubov coefficient for any $\lam \sqrt{c}/\kappa$, and allows a simple comparison to the Hawking result~\cite{Macher09b}.
An alternative way to define the temperature of a black hole is through the ratio $|\beta/\alpha|^2$, that must follow the Boltzmann law, that is
\be \label{DetailedBalance}
|\beta/\alpha|^2 = \exp(-\om/T_H).
\ee
This condition on the ratio $|\beta/\alpha|^2$ corresponds to the detailed balance condition. One of its virtue is that it is independent of greybody factors~\cite{Linder15} (although they are neglected in the KdV model, we point out that the notion of effective temperature would generalize more easily using \eqref{DetailedBalance}). Using the norm conservation $|\alpha|^2 - |\beta|^2 = 1$, we see that
\be
\left|\frac{\beta}{\alpha}\right|^2 = 1 - \frac1{|\beta|^2} \simeq 1 - \frac{\om}{T_\eff} \simeq \exp(- \om/T_\eff).
\ee
Our definition of $T_\eff$ coincide at low frequencies with the usual one, whether one considers $\alpha$ and $\beta$ separately, or their ratio. We shall now describe the behavior of the effective Hawking temperature in an exactly solvable example.

\subsubsection{Exactly solvable example}
Next we would like to calculate the effective Hawking temperature for the following profile
\be \label{exactWH}
c(x)-v(x) = c_r - v_r - \frac{c_r - v_r + v_l - c_l}{1+e^{x/L}}.
\ee
In this model, the analogue horizon (where $v = c$) is located at the point
\be
x_H = L \ln\left(\frac{v_l - c_l}{c_r - v_r}\right).
\ee
The surface gravity~\eqref{eq-surface-gravity} is then
\be
\kappa = \frac{(v_l - c_l) (c_r - v_r)}{L (v_l - c_l + c_r - v_r)}.
\ee
The general solution to \eq{SchroEq} is given in terms of hypergeometric functions, by
\bea
\conj &=& \mu_1 e^{-q_l x/L} {}_2 F_1\left(- q_l - iq_r, - q_l + iq_r; 1 - 2q_l; -e^{x/L}\right) , \nonumber \\
&& + \mu_2 e^{q_l x/L} {}_2 F_1\left(q_l + iq_r, q_l - iq_r; 1 + 2q_l; -e^{x/L}\right).
\eea
The bounded mode is obtained by setting $\mu_1 = 0$. The asymptotic of this solution is obtained by using transformation of variables for hypergeometric functions (see App.~\ref{Hyper_App}, \eq{TransVar}). We obtain
\bsub \bea
A_2 &=& \frac{\Gamma(1+2q_l) \Gamma(2i q_r)}{\Gamma(q_l + i q_r) \Gamma(1+ q_l + iq_r)}, \\
A_3 &=& \frac{\Gamma(1+2q_l) \Gamma(-2i q_r)}{\Gamma(q_l - i q_r) \Gamma(1+ q_l - iq_r)} .
\eea \esub
To obtain the constant term $A_1$, we integrate our solution $\conj$ along the flow, i.e. $A_1 = \int_{\mathbb R} \conj(x)dx$. Using another hypergeometric identity (see App.~\ref{Hyper_App}, \eq{TransVar}), we obtain
\be
A_1 = \frac{\pi L \Gamma(1+2q_l)}{q_l q_r \sinh(\pi q_r) |\Gamma(q_l+i q_r)|^2} .
\ee
We insert our findings for $A_1$ and $A_2$ in the general result for the $\beta$-coefficient~\eqref{eq-beta-general} in the low-frequency limit and obtain
\be \label{WHbeta}
|\beta|^2 = \frac{\kappa \sinh(\pi q_r)^2}{\pi \sinh(2\pi q_r) \om}. \bigskip
\ee

\textbf{Effective temperature for exactly solvable profiles.}
In the white hole flow described by \eq{exactWH}, the effective temperature is, from \eqref{WHbeta}, given by
\be \label{Teff}
T_\eff = \frac{\kappa \sinh(\pi q_r)^2}{\pi \sinh(2\pi q_r)} = \frac{\kappa}{2\pi} \tanh(\pi q_r).
\ee
Interestingly, this expression was conjectured in~\cite{RobertsonPhD} in the same profile as \eq{exactWH}, but in a model including the downstream mode. We have shown here that this expression is exact for the Korteweg-de Vries model in our exactly solvable profile of \eq{exactWH}.

The effective temperature \eqref{Teff} interpolates between two remarkable regimes. First the Hawking regime, or smooth regime, when $\lam L \sqrt{c} \to \infty$. We see that our expression becomes
\be
T_\eff = \frac{\kappa}{2\pi} \left(1+ O(e^{-2\pi q_r})\right),
\ee
which is the standard Hawking temperature. Remarkably, in this profile, the Hawking temperature is extremely robust, since we see that dispersive corrections are exponentially suppressed. The other limit is the limit of a discontinuous step profile, when $\lam L \sqrt{c} \to 0$. In such a case, one finds
\be
T_\eff \sim \frac{\kappa q_r}{2} =: T_{\rm step},
\ee
which coincide with what was found in the literature~\cite{Finazzi12,RobertsonPhD,Mayoral10}. Our general result allows us to see explicitly how the $\beta$ coefficient interpolates between the Hawking result and the step result. For instance, when decreasing $L$, $\kappa$ increases, and the temperature as well. But when $\lam L \sqrt{c} \sim 1$, it saturates to a maximum value $T_{\rm step}$. This is illustrated in Fig.~\ref{Teff_Fig}.

\begin{figure}[!ht]
\begin{center}
\includegraphics[width=0.6\columnwidth]{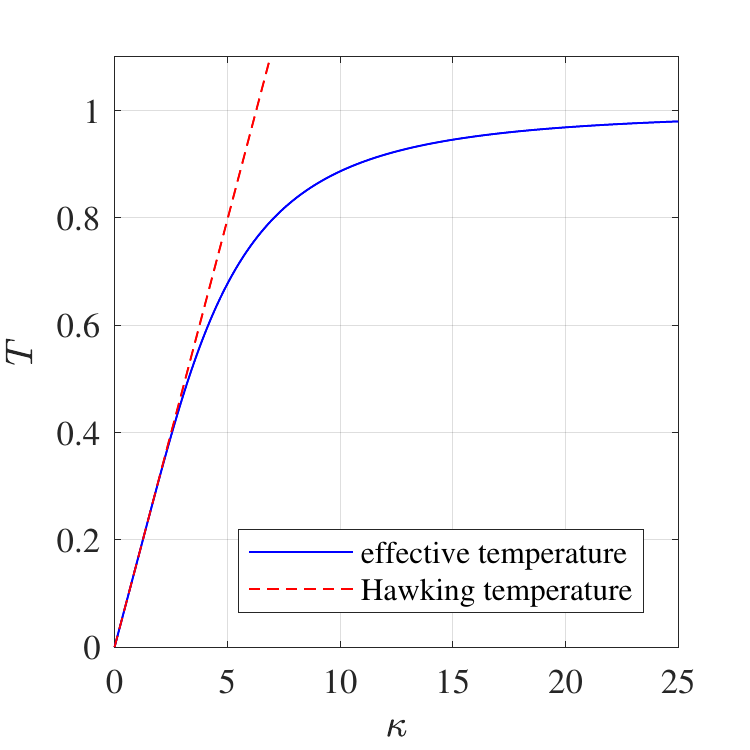}
\end{center}
\caption{Evolution of $T_\eff$ (solid blue line) of \eq{Teff} as a function of the surface gravity $\kappa$. The parameters of the flow ($v_{l/r}$ and $c_{l/r}$) and the dispersive scale $\lam$ are fixed, and such that $T_{\rm step} = 1$. The (red) dotted line is the Hawking temperature, given by $\kappa/2\pi$.}
\label{Teff_Fig}
\end{figure}

%%%%%%%%%%%%%%%%%%%%%%%%%%%%%%%%%%%%%%%%%%%%%%%%%%%
%SUBCRITICAL KdV
%%%%%%%%%%%%%%%%%%%%%%%%%%%%%%%%%%%%%%%%%%%%%%%%%%%

\subsection{Subcritical flows\label{Sec-Subcritical-flows}}
We now consider a subcritical flow in the KdV model with a subcritical dispersion relation. The other case is then obtain by symmetry (see \eq{S_sym}).

\subsubsection{General results}
Because the flow is now subcritical on both sides, solutions of \eq{SchroEq} are oscillating on both sides. This means that they encode dispersive modes present on both sides. Hence, it is easy to build the modes $\phi_\+^{\rm in}$ and $\phi_\-^{\rm in}$ with the same method as before. For instance, let's consider $\phi_\-^{\rm in}$. We integrate the relevant solution of \eqref{SchroEq}, such that the constant term is on the left side, that is
\bsub \bea
\phi_1 &\underset{-\infty}\sim& - \tilde A_1 + \frac{e^{i q_l x/L}}{i q_l /L}, \\
&\underset{+\infty}\sim& \tilde A_2 \frac{e^{i q_r x/L}}{i q_r/L} + \tilde A_3 \frac{e^{-i q_r x/L}}{-iq_r/L}.
\eea \esub
After normalization, this gives us the mode
\bsub \bea
\phi_\-^{\rm in} &\underset{-\infty}\sim& \frac1{\sqrt{\om}} \tilde \beta + \frac{e^{i q_l x/L}}{\sqrt{2(c_l - v_l) q_l/L}}, \\
&\underset{+\infty}\sim& \tilde A \frac{e^{i q_r x/L}}{\sqrt{2(c_r - v_r) q_r/L}} + \tilde B \frac{e^{-i q_r x/L}}{\sqrt{2(c_r - v_r) q_r/L}}.
\eea \esub
From this we extract the scattering coefficients, and in particular
\be
\tilde \beta = i \tilde A_1 \sqrt{\frac{q_l \om}{2(v_l - c_l) L}}.
\ee
We see that the general behavior of $|\tilde \beta|^2$ is linear in $\om$. This contrasts with the case of a transcritical flow, and agrees with the results of~\cite{Michel14,Coutant16,Robertson16}. Before analyzing this behavior in more details, we construct the mode $\phi_\u^{\rm in}$, schematically represented in Fig.~\ref{Vancouver_Fig}. The reason is that this modes was the one considered in the scattering experiments using water waves~\cite{Weinfurtner10,Euve14}. However, it is harder to obtain with the above method, since no dispersive mode is present on the left. To circumvent this difficulty, we focus on the $\beta$ coefficient, and build an \emph{out} mode instead: $\phi_\-^{\rm out}$. We then obtain $\beta$ by inverting the $S$-matrix \eqref{SubcritSmat}. Since $S \in U(1,2)$, its inverse is fairly simple, and given by
\be
S^{-1} = \bmat T^* & -\beta^* & \alpha^* \\ -\tilde \beta^* & \tilde A^* & -\tilde B^* \\ \tilde \alpha^* & -B^* & A^* \emat.
\ee
To obtain the \emph{out} mode $\phi_\-^{\rm out}$, we build the solution of \eqref{SchroEq} with the general asymptotic behavior
\bsub \label{phi_-_out} \bea
\phi_2
&\underset{-\infty}\sim& A_2 \frac{e^{i q_l x/L}}{i q_l /L} + A_3 \frac{e^{-i q_l x/L}}{-i q_l /L} , \\
&\underset{+\infty}\sim& A_1 + \frac{e^{-i q_r x/L}}{-i q_r /L} .
\eea \esub
Moreover, for $x \ll c/\om$, the mode $\phi_\-^{\rm out}$ has the following behavior
\bsub \bea
\phi_\-^{\rm out}
&\underset{-\infty}\sim& - B^* \frac{e^{i q_l x/L}}{\sqrt{2(c_l - v_l) q_l/L}} + \tilde A^* \frac{e^{-i q_l x/L}}{\sqrt{2(c_l - v_l) q_l/L}} , \\
&\underset{+\infty}\sim& -\frac{\beta^*}{\sqrt{\om}} + \frac{e^{-i q_r x/L}}{\sqrt{2(c_r - v_r) q_r/L}} .
\eea \esub
Identifying the two, we deduce the coefficient
\be \label{eq-beta-subcrtical}
\beta = -i A_1 \sqrt{\frac{\om q_r}{2(c_r - v_r) L}}.
\ee
As we see, it also vanishes at low $\om$ as $|\beta|^2 = O(\om)$. For later discussions on the behavior of $\beta$, it is convenient to define the characteristic frequency 
\be
\sigma = \lim_{\om \to 0} \left(\frac{\om}{|\beta_\om|^2}\right).
\ee
To obtain $\alpha$, the other Bogoliubov coefficient of the mode in Fig.~\ref{Vancouver_Fig}, we must build $\phi_+^{\rm out}$. However, it is easy to see that it directly follows from the complex conjugate of \eq{phi_-_out}, and hence $\alpha = \beta^*$. In fact, this comes from a more general property of the equation (see e.g.~\cite{Coutant16}, equation (40)). We now study the characteristic frequency $\sigma$ in the exactly solvable example, which as we saw, characterizes both $\alpha$ and $\beta$ (and a similar analysis would follow for the other scattering coefficients of \eq{SubcritSmat}).
\begin{figure}[!ht]
\begin{center}
\includegraphics[width=0.85\columnwidth]{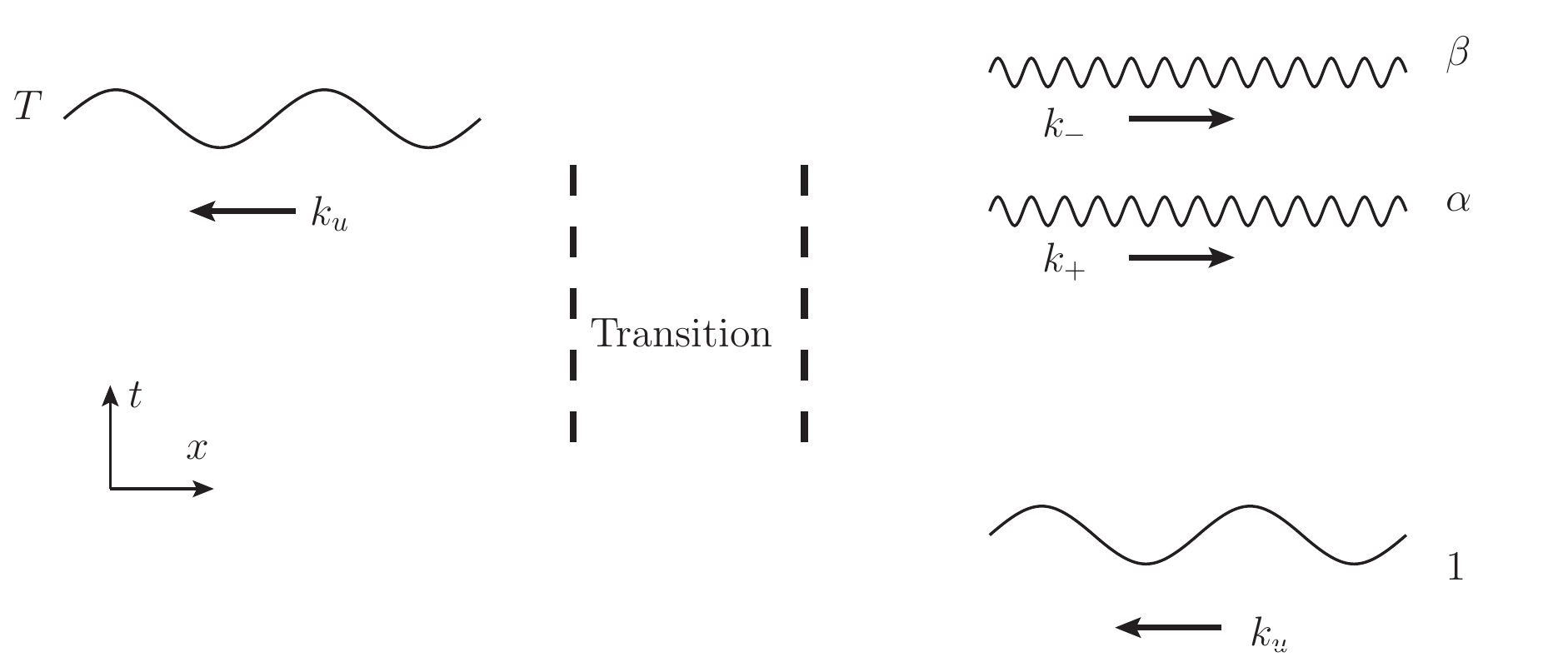}
\end{center}
\caption{Schematic representation of the mode $\phi_\u^{\rm in}$ on an accelerating subcritical flow.
}
\label{Vancouver_Fig}
\end{figure}

\subsubsection{Exactly solvable example}
We consider the same profile as in \eqref{exactWH}, but that stays subcritical, i.e. $0 < c_l - v_l < c_r - v_r$. We can use the same general solution as before with the replacement $q_l \to -i q_l$, that is
\bea
\conj &=& \mu_1 e^{-i q_l x/L} {}_2 F_1\left(-i q_l + iq_r, -i q_l - iq_r; 1 - 2 i q_l; -e^{x/L}\right) , \nonumber \\
&& + \mu_2 e^{i q_l x/L} {}_2 F_1\left(i q_l - iq_r, i q_l + iq_r; 1 + 2 i q_l; -e^{x/L}\right) .
\eea
To obtain the mode $\phi_\-^{\rm out}$, we use a transformation of variable, so as to control its behavior on the left side. The corresponding solution of \eq{SchroEq} is
\be
\conj = e^{- i q_r x/L} {}_2 F_1\left(i q_l + iq_r, - i q_l + iq_r; 1 + 2 i q_r; -e^{-x/L}\right),
\ee
Now, using
\be
A_1 = \frac{L \pi \Gamma(1 + 2 i q_r)}{i q_r q_l \sinh(\pi q_l) \Gamma(i q_l + i q_r) \Gamma(-i q_l + i q_r)},
\ee
and the preceding result, we obtain
\be
\sigma = \frac{(c_r - v_r) q_l^2}{\pi L (q_r^2 - q_l^2)} \frac{\sinh(\pi q_l)^2 \sinh(2\pi q_r)}{\sinh(\pi q_r - \pi q_l) \sinh(\pi q_r + \pi q_l)}.
\ee
Once again, it is instructive to discuss its various asymptotic behaviors. In the smooth limit $\Lam L \sqrt{c} \to \infty$, we have
\be \label{SCsigma}
\sigma \sim \frac{(c_r - v_r) q_l^2}{2\pi L (q_r^2 - q_l^2)} e^{2\pi q_l} = \frac{(c_r - v_r) (c_l - v_l)}{2\pi L (c_r - c_l + v_l - v_r)} e^{2\pi q_l}.
\ee
In particular, we see that contrary to the transcritical case, the mode mixing between positive and negative norm modes is driven by dispersion. Indeed, $\sigma$ increases (hence $\beta$ decreases) exponentially when $\lam \to \infty$. This result coincide with what was obtained using Bremmer series at low frequencies~\cite{Coutant16}. In the step limit $\lam L \sqrt{c} \to 0$, one obtains
\be
\sigma \sim \frac{2(c_r - v_r) q_l^4 q_r}{L (q_r^2 - q_l^2)^2} = \frac{\Lam (c_l - v_l)^2 (2(c_r - v_r))^{3/2}}{(c_r - c_l + v_l - v_r)^2} .
\ee
Another interesting limit is the near critical case, that is $q_l \to 0$, which gives
\be
\sigma \sim \frac{\pi (c_r - v_r) q_l^4}{L q_r^2} \frac{\sinh(2\pi q_r)}{\sinh(\pi q_r)^2}.
\ee
As we see, the power law in $q_l$, and hence in $(c_l - v_l)$ does not follow from the semiclassical formula (\eq{SCsigma}), which is obtained for $q_l \gg 1$. However, the semiclassical formula can still be quite accurate for near-critical flows $|1 - v_l/c_l | \ll 1$, as long as $|1 - v_l/c_l | \gg 1/(\lam L \sqrt{c})$, which is consistent since $1/(\lam L \sqrt{c}) \ll 1$ in the semiclassical limit. Roughly speaking, the semiclassical formula is valid for near critical flows, but not extremely near critical flows.

%%%%%%%%%%%%%%%%%%%%%%%%%%%%%%%%%%
% CONCLUSION
%%%%%%%%%%%%%%%%%%%%%%%%%%%%%%%%%%
\section{Conclusion}

We studied the Hawking radiation and its modifications due to dispersive effects using the linearized Korteweg-de Vries model. We first motivate in details the usefullness of the Korteweg-de Vries model. It allows us to disentangle the modifications of the Planck spectrum between those due to greybody factors, and those induced by dispersion. We argue that this model is accurate for near critical flows, and by compassion in the companion paper~\cite{Coutant17b}, we demonstrate this statement at the level of the $S$-matrix. The Korteweg-de Vries model is quite popular for surface waves, but as we argue it also provides a good approximation of the analogue Hawking effect in Bose-Einstein condensates. In Appendix~\ref{KdV_App}, we give an alternative derivation, starting from the effective metric.
\bigskip

Within the Korteweg-de Vries model, one sees that there is not only a correspondence between black and white hole flows by time reversal, but also a similar correspondence between the two types of dispersion relations: whether it decreases the propagation speed as shorter wavelengths (subcritical dispersion), or whether it increases (supercritical), see \eq{S_sym}. This generalizes a result obtained in~\cite{Coutant11}. Hence, one needs to solve the scattering problem only once to obtain the result for the other cases.
\bigskip

We then develop a matched asymptotic expansion method to solve the scattering problem in the low-frequency limit. This method allows us to demonstrate a number of results that were observed numerically~\cite{Finazzi11,Finazzi12,RobertsonPhD,Robertson16}, and generalize them. First we show that the Bogoliubov coefficients increase as the inverse square root of the frequency when the flow is transcritical (see \eq{eq-beta-general}). On the other hand, if the flow is subcritical (resp. supercritical) all along, and the dispersion relation subcritical (resp. supercritical), the Bogoliubov coefficients vanish as the square root of the frequency (see \eq{eq-beta-subcrtical}).
\bigskip

We then apply it to an exactly solvable flow profile. For transcritical flows, the effective temperature (defined as the low-frequency limit of $\om |\beta_\om|^2$) interpolates between the Hawking temperature, when the flow gradients are smaller than the scale of dispersion, to the result of a step profile, when gradients are bigger than the dispersive scale. In this profile, we found that deviations to the Hawking temperature are exponentially small, rather than polynomial. This explains why the agreement between dispersive theories and the relativistic wave equation is excellent when greybody factors are small~\cite{Macher09b,Finazzi12}. In flows that stay subcritical or supercritical, we also derive exact expressions. As was obtained in~\cite{Coutant16}, the Bogoliubov coefficients originate from dispersion, and decrease exponentially in the non-dispersive limit.

\acknowledgements
We would like to thank Renaud Parentani and Scott Robertson for useful comments about the final version of this manuscript. This project has received funding from the European Union's Horizon 2020 research and innovation programme under the Marie Sk\l odowska-Curie grant agreement No 655524. S.~W.~acknowledges financial support provided under the Royal Society University Research Fellow (UF120112), the Nottingham Advanced Research Fellow (A2RHS2), the Royal Society Project (RG130377) grants and the EPSRC Project Grant (EP/P00637X/1).

%%%%%%%%%%%%%%%%%%%%%%%%%%%%%%%%%%
% APPENDIX
%%%%%%%%%%%%%%%%%%%%%%%%%%%%%%%%%%
\newpage
\appendix
%%%%%%%%%%%%%%%%%%%%%%%%%%%%%%%%%%
% CONFORMAL WAVE AND KDV
%%%%%%%%%%%%%%%%%%%%%%%%%%%%%%%%%%
\section{From the wave equation in curved space-time to the linearized Korteweg-de Vries equation}
\label{KdV_App}
We argue here that the linearized Korteweg-de Vries equation correctly describes the propagation of modes propagating against the flow. For this we need 2 conditions to be realized. First, we must be in a ``weak dispersive regime''; that is, dispersive effects must be weak enough so that it is sufficient to consider only the first non-relativistic corrections. Second, modes propagating with the flow must decouple from the one propagating against it. In the gravitational language, greybody factors must be negligible. Due to conformal invariance, a scalar field in a 1+1 dimensional metric automatically satisfies this decoupling condition. Therefore we start by considering the 1+1 space-time metric described by the line element
\be \label{ds2}
ds^2 = c^2 dt^2 - (dx - v dt)^2,
\ee
where $c(x)$ is the local wave speed, and $v(x)$ the velocity of the background flow. In this space-time, the wave equation reads
\be \label{Conf_eq}
(\p_t + \p_x v) \frac{1}{c} (\p_t + v\p_x) \phi - \p_x c \p_x \phi = 0.
\ee
Conformal invariance of this equation implies that every solutions of \eq{Conf_eq} decomposes as $\phi = \phi_\u + \phi_\v$, where $\phi_\u$ and $\phi_\v$ obey the set of (uncoupled) first order equations
\bsub \label{Split_eq} \bea
(\p_t + v\p_x) \phi_\u &=& c \p_x \phi_\u, \label{Conf_u} \\
(\p_t + v\p_x) \phi_\v &=& - c \p_x \phi_\v \label{Conf_v}.
\eea \esub
For $v>0$, the subscript $\u$ refers to counter-propagating modes, while $\v$ refers to co-propagating modes. There are now two ways of introducing short wavelength dispersion to this equation. The first is to add higher spatial derivatives to the full \eq{Conf_eq}. This was the historical choice, initiated by Jacobson~\cite{Jacobson95} and Unruh~\cite{Unruh95}. One disadvantage of this choice is that it breaks the exact decoupling between the $\u$ and $\v$ sectors. That is, the dispersive equation do not split as in \eq{Split_eq}. The second option is to introduce dispersion separately for the $\u$ and $\v$ sectors, i.e.
\bsub \bea
(\p_t + v\p_x) \phi_\u &=& c \p_x \phi_\u + f(\p_x) \phi_\u,  \\
(\p_t + v\p_x) \phi_\v &=& - c \p_x \phi_\v - f(\p_x) \phi_\v.
\eea \esub
This second option was first proposed in~\cite{Schutzhold08}. Now, if we restrict ourselves to the weak dispersive regime, the first correction that preserves parity invariance is
\be \label{disp_function}
f(\p_x) = \pm \p_x^3/\lam^2.
\ee
Therefore, the propagation equations become
\bsub \label{uv_eq} \bea
(\p_t + v\p_x) \phi_\u &=& c \p_x \phi_\u \pm \p_x^3/\lam^2 \phi_\u, \label{u_eq} \\
(\p_t + v\p_x) \phi_\v &=& - c \p_x \phi_\v \mp \p_x^3/\lam^2 \phi_\v. \label{v_eq}
\eea \esub
When we restrict ourselves to $u$-modes, i.e. \eq{u_eq}, we obtain the linearized Korteweg-de Vries equation used in the core of the text. In general, perturbations of a one-dimensional flow do not obey the 1+1 dimensional wave equation, and hence, the $\u$ and $\v$ sectors are not decoupled. In our companion work~\cite{Coutant17b},  we study in detail this coupling at low frequencies (for which it is most significant), and under what conditions it can be neglected. In particular, we show that in the near critical limit ($|1-v/c| \ll 1$), both $u$-$v$ decoupling and weak dispersion are a good approximation.

In a transcritical flow, i.e. $v$ crosses $c$ once, the set of equation \eqref{uv_eq} can describe 4 different cases: waves propagating on a black hole flow ($v$ accelerates) or a white hole flow ($v$ decelerates), both with two types of dispersion (choice of sign in \eq{disp_function}). In fact, it is enough to compute the $S$-matrix for one case, to obtain it for the 3 others, using discrete symmetries of the system~\cite{Coutant11}. First, when focusing on the counter-propagating sector \eqref{u_eq}, we see that it is equivalent to change the sign of the dispersive term, or to perform the change
\bsub \bea
t &\to& -t , \\
c-v &\to& -(c-v) .
\eea \esub
The first line changes the time direction, and therefore, exchanges the role of \emph{in} and \emph{out} modes. As a result, the $S$-matrix is changed into its inverse $S^{-1}$. The second line exchanges the supersonic and subsonic sides. In particular, a black hole flow becomes a white hole one, or \emph{vice-versa}. In addition, one has the usual correspondence between black and white hole by time reversal~\cite{Macher09b}. More precisely, swapping the roles of \eqref{u_eq} and \eqref{v_eq} is equivalent to the change
\bsub \bea
t &\to& -t , \\
v &\to& -v .
\eea \esub
More simply, one goes from the $S$-matrix of a black hole flow to the one of a white hole flow by taking its inverse. Using these correspondences, we obtain the relations between the 4 cases, as used in the text,
\be
S_{\rm BH}^+ = (S_{\rm WH}^+)^{-1} = (S_{\rm WH}^-)_{l \leftrightarrow r}^{-1} = (S_{\rm BH}^-)_{l \leftrightarrow r}. \nonumber
\ee

%%%%%%%%%%%%%%%%%%%%%%%%%%%%%%%%%%
% HYPERGEOM FORMULAS
%%%%%%%%%%%%%%%%%%%%%%%%%%%%%%%%%%
\section{Useful properties of hypergeometric functions}
\label{Hyper_App}
In this paper, following references~\cite{Olver,DLMF} we defined hypergeometric functions as
\be
{}_2 F_1(a,b;c;z) = \sum_{n=0}^\infty \frac{(a)_n (b)_n}{(c)_n n!} z^n = \sum_{n=0}^\infty \frac{\Gamma(a+n) \Gamma(b+n) \Gamma(c)}{\Gamma(a) \Gamma(b) \Gamma(c+n) n!} z^n.
\ee
In the text, solutions of the second order differential equation are given in terms of hypergeometric functions. To obtain their asymptotic behavior, we use the transformations of variables
\bea \label{TransVar}
{}_2 F_1(a,b;c;z) &=& \frac{\Gamma(c) \Gamma(b-a)}{\Gamma(b) \Gamma(c-a)} (-z)^{-a} {}_2 F_1(a,a-c+1;a-b+1; z^{-1}) , \nonumber \\
&&+ \frac{\Gamma(c) \Gamma(a-b)}{\Gamma(a) \Gamma(c-b)} (-z)^{-b} {}_2 F_1(b,b-c+1;b-a+1; z^{-1}).
\eea
Since $z\to 0$ and $z \to \infty$ are the two asymptotic regions of the scattering problem, the above transformation allows us to extract scattering coefficients, as done in the core of the paper. Moreover, we also need to integrate these solutions to obtain the constant contribution accumulated from one side to the other. For this, we need another useful identity~\cite{DLMF}:
\be
\int_0^{\infty} y^{d-1} {}_2 F_1(a,b;c;-y) dy = \frac{\Gamma(d) \Gamma(c) \Gamma(a-d) \Gamma(b-d)}{\Gamma(a) \Gamma(b) \Gamma(c-d)}.
\ee

\bibliographystyle{utphys}
\bibliography{Bibli}

\end{document}